\documentclass{myws-procs9x6}

\usepackage{mywsamsmath}
\usepackage{neb-x}
\usepackage{graphicx}

\begin{document}

\title{
Radiative fields in spacetimes with Minkowski and de~Sitter asymptotics%
\footnote{\uppercase{A}n extended version of the invited talk given by \uppercase{J}.~\uppercase{B}. at
the 10th \uppercase{G}reek \uppercase{R}elativity \uppercase{M}eeting
in \uppercase{K}allithea, \uppercase{C}halkidiki, \uppercase{M}ay 2002.
\uppercase{S}ecs.~\protect\ref{sec:intro}, \protect\ref{sec:flatex}, \protect\ref{sec:boostrot} and \protect\ref{sec:rtwl}
are in part based on the reviews quoted as
\uppercase{R}efs.~\protect\refcite{bicak:ehlers}, \protect\refcite{bicak:1997}
and \protect\refcite{bicak:2000}.}
}

\author{Ji\v{r}\'i Bi\v{c}\'ak and Pavel Krtou\v{s}}

\address{%
Institute of Theoretical Physics,\\
Faculty of Mathematics and Physics, Charles University,\\
V Hole\v{s}ovi\v{c}k\'{a}ch 2, 180 00 Prague 8, Czech Republic\\
}

\date{March 4, 2003} 

\maketitle

\abstracts{
The classical Bondi-Penrose approach to the gravitational radiation theory
in asymptotically flat spacetimes is recalled and recent advances in
the proofs of the existence of such spacetimes are briefly reviewed.
We then mention the unique role of the boost-rotation symmetric spacetimes,
representing uniformly accelerated objects, as the only explicit radiative
solutions known which are asymptotically flat; they are used as test
beds in numerical relativity and approximation methods.\newline
The main part of the review is devoted to the examples of radiative
fields in the vacuum spacetimes with positive cosmological constant.
Type {\it N} solutions are analyzed by using the equation of geodesic deviation.
Both these and Robinson-Trautman solutions of type {\it II} are
shown to approach de~Sitter universe asymptotically.
Recent work on the the radiative fields due to uniformly accelerated charges
in de~Sitter spacetime (``cosmological Born's solutions'') is reviewed and
the properties of these fields are discussed with a perspective to characterize
general features of radiative fields near a de~Sitter-like infinity.
}

\section{Introduction}
\label{sec:intro}

After several important contributions to the gravitational radiation theory in
the late 1950's and early 1960's by Pirani, Bondi, Robinson, Trautman and
others, a landmark paper by Bondi et al\cite {BondiBurgMetzner:1962} appeared
in which radiative properties of isolated (spatially bounded) axisymmetric
systems were studied along outgoing null hypersurfaces ${u=\text{constant}}$,
with $u$ representing a retarded time function. An ansatz was made that the
asymptotically Minkowski metric along ${u=\text{constant}}$ can be expanded in
inverse powers of~$r$,
\begin{equation}\label{RSE1}
g_{\mu v} = \eta_{\mu v}+h_{\mu v}(\theta) r^{-1}+f_{\mu v}(\theta)r^{-2}+\ldots,
\end{equation}
where $r$ denotes a suitable parameter along null generators (parametrized by
coordinates $\theta, \varphi$) on the hypersurfaces ${u=\text{constant}}$. Under the
assumption (\ref{RSE1}), Einstein's vacuum equations with the vanishing cosmological constant
$\Lambda$ were shown to determine
uniquely formal power series solution of the form (\ref{RSE1}), provided that a
free "news function" $c(u, \theta)$ is specified. The news function contains
all information about radiation at infinity $("r=\infty")$. It enters the
fundamental "Bondi mass-loss formula" for the total mass $M(u)$ of an isolated
system at retarded time $u$. Field equations imply that $M(u)$ is a
monotonically decreasing function of $u$ if $\partial_u c \not= 0.$ A natural
interpretation is that gravitational waves carry away positive energy from the
system and thus decrease its mass. In the work of Bondi et al\cite{BondiBurgMetzner:1962} as well
as in the important generalizations by Sachs, Newman and Penrose (see, e.g., Ref.~\refcite{NewmanTod:1980}), the decay of
radiative fields was studied in preferred coordinate systems.

In 1963 Penrose\cite{Penrose:1963} formulated a beautiful {\it geometrical} framework for
description of the "radiation zone" in general relativity in terms of conformal
infinity.

Penrose's definition of asymptotically flat radiative spacetimes avoids such
problems as "distances" or "suitable coordinates", and incorporates a clear
definition of what is infinity. It is inspired by the work on radiation theory
mentioned above, and by the properties of conformal infinity in Minkowski
spacetime. In contrast to an Euclidean space, one can go
to infinity in various directions: moving along timelike geodesics we come to
the future (or past) timelike infinity $I^+$ (or $I^-$); along null geodesics
(cf. Eq. (\ref{RSE1})) we reach the future (past) null infinity ${\scri}^+
({\scri}^-)$; spacelike geodesics lead to spatial infinity ${\it i}_0$.
An asymptotically flat spacetime can be compactified and mapped into a finite region by an
appropriate conformal transformation. Thus one obtains the well-known
Penrose diagram in which the three types of infinities are mapped into the
boundaries of the compactified spacetime.

It is generally accepted that Penrose's definition forms the only rigorous,
geometrical basis for the discussion of gravitational radiation from isolated
systems. It enables us to use techniques of local geometry "at infinity".
For example, in the classical papers by Bondi et al\cite{BondiBurgMetzner:1962,NewmanTod:1980},
the decay of the curvature along outgoing null geodesics at infinity exhibits "the
peeling-off" properties: the fall-off of various components of the Weyl tensor is
related to their Petrov algebraic type. To be more specific, certain complex
linear combinations of the Weyl tensor in the orthonormal frame, $\Psi_k
(k=0,1,2,3,4)$, behave as $\Psi_k = O (r^{k-5})$ as $r \rightarrow \infty$. (In
particular, $\Psi_4 \sim r^{-1}$ has the same algebraic structure as the Weyl
tensor of a plane wave -- radiative field of a bounded system resembles
asymptotically that of a plane wave.)  This decay of the curvature can be shown
to follow (see e.g.\cite{Geroch:1997}) from a sufficient differentiability (smoothness)
of the conformally rescaled (unphysical) metric $\tilde g$
at the boundary representing null infinity.

In the next section we shall review some delicate rigorous mathematical
results proving the existence of asymptotically flat vacuum spacetimes.
We shall see that although such spacetimes satisfying the peeling properties
have recently been shown to exist, it is far from clear what is their status in
completely generic situations, in particular when sources are present.
The boost-rotation symmetric spacetimes, representing uniformly accelerated
``particles'' or black holes, the role of which will be briefly discussed in
Sec.~\ref{sec:boostrot}, do exhibit the peeling properties,\cite{Bicak:Robinson} but
they contain two Killing vectors, and although they admit global smooth null infinity,
this is singular at the points where particles ``start'' and ``end''.

Surprisingly perhaps, the existence of generic spacetimes was proved more then
15 years ago in the case of vacuum spacetimes with a positive cosmological
constant\cite{Friedrich:1986,Friedrich:1998} (Sec.~\ref{sec:dSex}).
However, until now a little attention, within the framework of exact theory,
has been paid to gaining a more ``physical'' picture of radiation
propagating in spacetimes which are not asymptotically flat.
In Secs.~\ref{sec:nwl} and \ref{sec:rtwl} we discuss exact solutions of Petrov types
{\it N} and {\it II} of the vacuum field equations with $\Lambda>0$
and interpret them as waves which at large times --- near future infinity which is in
this case spacelike --- decay and leave ``bald'' de~Sitter spacetime
(cosmic no-hair). Finally, in Sec.~\ref{sec:fieldsinds}, which is more technical
and detailed than preceeding sections, we analyze the scalar and
electromagnetic test fields from uniformly accelerated charges in
de~Sitter spacetime. And in the concluding ``Outlook'', we comment on
more general cases of (non-test) fields near a spacelike infinity of de~Sitter type.

It is natural to use de~Sitter space for studying radiating sources in
spacetimes which are not asymptotically flat and possess disjoint future and past infinities
which are spacelike. It is the space of constant curvature, conformal to Minkowski space, and with
Huygens principle satisfied for conformally invariant fields. The de~Sitter
universe also plays an important role in cosmology --- not only in the context
of inflationary theories but also as the \vague{asymptotic state} of standard
cosmological models with $\Lambda>0$, which has been indeed suggested by recent
observations.

\section{Asymptotically flat radiative spacetimes: existence}
\label{sec:flatex}

Despite its rigour and elegance, Penrose's definition of asymptotically flat spacetimes might turn to be of
a limited importance if no interesting radiative spacetimes {\it exist} which
satisfy the definition. In Section \ref{sec:boostrot} we shall describe special exact radiative
spacetimes which represent "uniformly accelerated" sources in general
relativity and admit $\scri$ as required; however, at least four points on
$\scri$ - those in which worldlines of the sources start and end - are
singular. There are no other {\it explicit} exact radiative solutions
describing finite sources available at present and this situation will probably
not change soon. Nevertheless, thanks to the work of Friedrich\cite{Friedrich:1998},
Christodoulou and Klainerman\cite{ChristodoulouKlainerman:book}, and most recently due to works of
Corvino and Schoen\cite{Corvino:2000,CorvinoSchoen:2003} and Chru\'{s}ciel and Delay\cite{ChruscielDelay:2002}
we know that globally non-singular
(including {\it all} $\scri$ ) asymptotically flat exact vacuum solutions of
Einstein's equations really exist.

The key idea of Friedrich is in realizing that Penrose's treatment of infinity
not only permits to use the methods of local differential geometry at $\scri$
but also to analyze global existence problems of solutions of Einstein's
equations in the physical spacetime by solving initial value problems in
the conformally related unphysical compact spacetime.
By using this approach Friedrich established that formal Bondi-type expansions
(\ref{RSE1}) converge locally at ${\scri}^+$. He succeeded to show that one can
formulate the "hyperboloidal initial value problem" for Einstein's vacuum
equations in which initial data are given on a hyperboloidal spacelike
hypersurface $\mathcal H$ which intersects $\scri^+$. It can then be
proven that hyperboloidal initial data, which are sufficiently close to
Minkowskian hyperboloidal data (i.e. to the metric induced on the hypersurface
$\mathcal H$ in Minkowski spacetime by the standard Minkowski metric), evolve to
a vacuum spacetime which is smooth on $\scri^+$ and ${\it I}^+$ as required by
Penrose's definition.

However, in a ``complete picture'' we would like to have initial data given on a {\it
standard} spacelike Cauchy hypersurface which does not intersect $\scri^+$ but
"ends" at spatial infinity, rather
than data given on a hyperboloidal initial hypersurface.
A remarkable progress in proving rigorously the existence of general,
asymptotically flat radiative spacetimes was achieved by Christodoulou and
Klainerman. Their treatise\cite{ChristodoulouKlainerman:book} contains the first really global general
existence statement for full, nonlinear Einstein's vacuum equations with
vanishing cosmological constant: Any smooth asymptotically flat initial data
set (determined by the first and the second fundamental form on a Cauchy
hypersurface) which is "near flat (Minkowski) data" leads to a unique, smooth
and geodesic complete development solution of Einstein's vacuum equations. This
solution is "globally asymptotically flat" in the sense that the curvature
tensor decays to zero at infinity in all directions.

The work of Friedrich, Christodoulou, Klainerman and others demonstrates
rigorously that the general picture of null infinity is compatible with the
vacuum Einstein field equations. However, important open questions remain.
As noted in Introduction, the
decay of the curvature (characterized by the Weyl tensor) along outgoing null
geodesics at infinity exhibits the peeling properties.
The results of Christodoulou and Klainerman, however, show a weaker peeling.
They were only able to prove that the asymptotically flat vacuum initial data
lead to $\Psi_0 \sim r^{-\frac{7}{2}}$ ({\it not} $\sim r^{-5}$) at null
infinity.

That {\it particular} radiative spacetimes exist
which are ``asymptotically simple'', i.e.,
they satisfy all Penrose's requirements, was
proved by Cutler and Wald\cite{CutlerWald:1989}. They show how to construct data on
a standard Cauchy hypersurface $\mathcal C$ such that the data are
near to flat spacetime data and coincide exactly with standard
vacuum Schwarzschild data outside of a compact region. Hence,
{\it on} initial hypersurface, there is a spherically symmetric
vacuum gravitational field outside the compact region. As a
consequence the future maximal evolution of the data is large
enough to contain a Friedrich's hyperboloidal hypersurface
$\mathcal H$ which
terminates on $\scri^+$. On this hypersurface the data evolved
from the original data will still be close to flat spacetime
data, i.e., to Minkowskian hyperboloidal initial data. By
involving Friedrich's theorem we then know that the spacetime
will be smoothly asymptotically flat in the future of this
hypersurface.
In fact, Cutler and Wald did not succeed in constructing the solutions
within the Einstein theory in vacuum but, rather, within the
Einstein-Maxwell theory.

Recently, however, Corvino and Schoen\cite{Corvino:2000} have shown
how vacuum Schwarzschild initial data can be glued to asymptotically
flat time-symmetric vacuum data in compact region. (For the most recent
generalization to the Kerr data see Ref.~\refcite{CorvinoSchoen:2003}.)
Starting from their construction Chru\'{s}ciel and Delay\cite{ChruscielDelay:2002}
obtained vacuum initial data which, using Friedrich's results,
lead to asymptotically simple radiative spacetimes. Hence, one knows at
present how to produce an infinite dimensional family of
vacuum, asymptotically flat radiative spacetimes which exhibit
the peeling properties and yield smooth $\scri$.

In a completely general situation one may still encounter so-called {\it
polyhomogeneous}\cite{Chruscieletal:1995} $\scri$, rather than smooth $\scri$.
The metric is called polyhomogeneous if at large {\it
r} it admits an expansion in terms of $r^{-j}\log^i r$ rather than $r^{-j}$ (as
it has been assumed in the works of Bondi and others - cf. Eq. (\ref{RSE1})).
The hypothesis of polyhomogeneity of $\scri$ has been shown to be formally
consistent with Einstein's vacuum equations, the
Bondi mass-loss law can be formulated, and the peeling properties of the
curvature hold, with the first two terms identical to the standard peeling,
the third term being $\sim r^{-3} \log r$.

In his more recent investigations\cite{Friedrich:1998}, Friedrich constructed the new -
finite but "wider" than the point $i_0$ - representation of spacelike infinity
which enables one to make much deeper analysis of the initial data
in the region where null infinity touches spacelike infinity. Good chances now
exist to obtain clear criteria determining which data lead to the smooth and
which just to the polyhomogeneous null infinity. First results in this direction
have been obtained for the linearized spin-2 equations\cite{Friedrich:2003}.

The ultimate goal of rigorous work on the existence and asymptotics of
solutions of the Einstein equations is {\it physics}: one hopes to be able to
consider (astro)physical \emph{sources}, to relate their behaviour to the
characteristics of the far fields. One would like to have under control various
(both analytical and numerical) approximation procedures. A still more
ambitious program is to consider strong initial data so as to be able to
analyze such issues as cosmic censorship.

\section{Asymptotically de~Sitter radiative spacetimes: existence}
\label{sec:dSex}

Curiously enough, in the case of vacuum Einstein's equations with a {\it
non-vanishing cosmological constant $\Lambda$} a more complete picture is known for some
time already. By using his regular conformal field equations, Friedrich
demonstrated\cite{Friedrich:1986} that initial data sufficiently close to de-Sitter data
develop into solutions of Einstein's equations with a positive cosmological
constant, which are asymptotically simple (with a smooth conformal infinity),
as required in the original framework of Penrose. In the case ${\Lambda>0}$
the infinity splits into two disjoint parts $\scri^+$ and $\scri^-$ which are both
\emph{spacelike}. Friedrich's theorem\cite{Friedrich:1986,Friedrich:1998} proves ``nonlinear stability of
asymptotic simplicity of de~Sitter space'': Changing the de~Sitter Cauchy data
by a finite, but sufficiently small amount, one gets again a solution satisfying
the definition of asymptotic simplicity for ${\Lambda>0}$. This involves
completeness of null geodesics, behavior of the conformal factor and metrics\cite{Friedrich:1986,Friedrich:1998}.
However, the directional dependence of radiative fields near
spacelike infinities has not been studied so far. This point will be discussed in
Secs.~\ref{sec:fieldsinds} and \ref{sec:outlook}.

Later Friedrich\cite{Friedrich:1995} also
analyzed the existence of asymptotically simple solutions to the Einstein
vacuum equations with a negative cosmological constant.

\section{Asymptotically flat radiative spacetimes with boost-rotation symmetry}
\label{sec:boostrot}

\begin{figure}[b]
\begin{center}
\includegraphics[width=7.5cm]{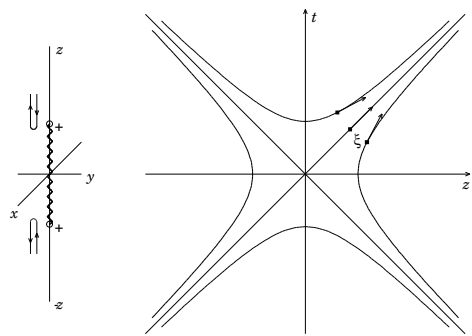}
\caption{\label{fig:brst}
Two particles uniformly accelerated in opposite directions.
The orbits of the boost Killing vector (thinner hyperbolas) are spacelike
in the region $t^2>z^2$.
Here, the boost-rotational symmetric metrics can \emph{locally} be transformed
into the metrics of Einstein-Rosen cylindrical waves.}
\end{center}
\end{figure}

These spacetimes representing ``uniformly accelerated objects'' (see
Fig. \ref{fig:brst}) have been reviewed in various places
(see e.g.\cite{bicak:ehlers,bicak:1997,bicak:2000,Pravdovi:2000} and references therein);
here we shall just mention few new developments.

The unique role of the boost-rotation symmetric spacetimes is exhibited by a
theorem (see Ref. \refcite{BicakPravdova:1998} and references therein) which roughly states that  in {axially} symmetric, locally
asymptotically flat electrovacuum spacetimes (in the sense that a null infinity
satisfying Penrose's requirements exists, but it need not necessarily exist
globally), the only {additional} symmetry that does not exclude radiation is
the {\it boost} symmetry.

In Ref.~\refcite{BicakPravdova:1998} the general functional forms of the news functions (both
gravitational and electromagnetic), and of the mass aspect and total Bondi mass
of boost-rotation symmetric spacetimes are given (see also Ref.~\refcite{Bicak:Robinson}). Recently similar results
were obtained\cite{Valiente-Kroon:2000} by using the Newman-Penrose formalism and under more
general assumptions (for example, {$\scri$} could in principle be
polyhomogeneous).

The general structure of the boost-rotation symmetric spacetimes with
hypersurface orthogonal Killing vectors was analyzed in detail in\cite{BicakSchmidt:1989}.
Their radiative properties, including explicit  construction of radiation
patterns and of Bondi mass for the specific boost-rotation symmetric solutions
were investigated in several works -- we refer to the reviews\cite{bicak:ehlers,bicak:1997,Bicak:Robinson,Pravdovi:2000}
for details. Recently the Newtonian limit of these spacetimes
was analyzed\cite{Valiente-KroonLazkoz:2002}.
The boost-rotation
symmetric spacetimes have also played the role
in such diverse fields like numerical relativity and
quantum production of black-hole pairs\cite{MannRoss:1995}.
For the most recent use of the boost-rotation symmetric solutions as test beds in numerical relativity,
see Ref.~\refcite{FrauendienerHein:2002}, where Friedrich's conformal field equations
are solved numerically with specific boost-rotation symmetric solutions being employed to check the code.

\section{On physical interpretation of non-twisting type {\it N} solutions with ${\Lambda\neq0}$}
\label{sec:nwl}

In order to gain an intuitive picture of radiative spacetimes with
${\Lambda\neq0}$, we considered\cite{BicakPodolsky:1999a,BicakPodolsky:1999b}
all non-twisting Petrov-type {\it N} solutions of vacuum Einstein's field equations
with cosmological constant. They belong either to the non-expanding Kundt class ${KN(\Lambda)}$
or to the expanding Robinson-Trautman class ${RTN(\Lambda)}$. We defined invariant subclasses of
each class and gave the corresponding metrics explicitly in suitable canonical
coordinates and interpreted them by analyzing the equation of geodesic
deviation.

In the case of the Kundt class, a suitable coordinate system
$(v,\xi,\bar\xi,u)$, where $\xi,\bar\xi$ are space-like coordinates, $v$ is a
parameter along the null geodesics tangent to the null vector $\bk$ and $u$ is a retarded time
with $u=\text{constant}$ being a wavefront, can be introduced in
which the $KN(\Lambda)$ metrics have the form
\begin{equation}
ds^2=2\frac{1}{p^2}d\xi d\bar\xi-2\frac{q^2}{p^2}dudv + F du^2\ , \label{E3.2}
\end{equation}
where
\begin{eqnarray}
&&p=1+ \la \xi\bar\xi\ , \qquad
q=(1-\la \xi\bar\xi)\alpha+\bar\beta\xi+\beta\bar\xi\ , \nonumber \\
&&F=\kappa{\frac{q^2}{p^2}}v^2 - {\frac{(q^2),_u}{p^2}}v - {\frac{q}{p}}H\ , \qquad
\kappa={\frac\Lambda3}\alpha^2+2\beta\bar\beta\ . \nonumber
\end{eqnarray}
Here $\alpha(u)$ and $\beta(u)$ are {\it arbitrary} real and complex functions
of $u$, respectively. These functions play the role of two arbitrary
`parameters', i.e., we can denote the Kundt class by $KN(\Lambda)\equiv
KN(\Lambda)[\alpha,\beta]$.
The function $H=H(\xi,\bar\xi,u)$ entering $F$ is
restricted by Einstein's equations, $H_{,\xi\bar\xi}+(\Lambda/3p^2)\,H=0$.
There exists a general solution to this equation
\begin{equation}
H(\xi,\bar\xi,u)=(f_{,\xi}+\bar f_{,\bar\xi})-{\frac\Lambda3p}
    (\bar\xi f + \xi\bar f)\ ,
\label{E3.5}
\end{equation}
where $f(\xi,u)$ is an arbitrary function of $\xi$ and $u$, analytic in $\xi$.
The space-time is conformally flat if and only if the structural function $H$
is of the form
\begin{equation}
H=H_c={\frac1p}{\Big[(1-\la \xi\bar\xi){\mathcal A}+
                       \bar {\mathcal B}\xi+{\mathcal B}\bar\xi\Big]}\ ,
\label{E3.6}
\end{equation}
with ${\mathcal A}(u)$ and ${\mathcal B}(u)$ being arbitrary real and complex
functions, respectively. Since $H_c$ of this form corresponds to (\ref{E3.5})
for $f$ quadratic in $\xi$ we easily infer that
the $KN(\Lambda)$ solutions (\ref{E3.2}), (\ref{E3.5})
with $f=f_c= c_0(u)+c_1(u)\xi+c_2(u)\xi^2$, where $c_i(u)$ are arbitrary
complex functions of $u$, are isometric to Minkowski (if $\Lambda=0$),
de~Sitter ($\Lambda>0$) and anti-de~Sitter ($\Lambda<0$) spacetime.

The Robinson-Trautman solutions\cite{RobinsonTrautman:1962} satisfying the vacuum
equations with $\Lambda$ can be written as\cite{Krameretal:book}
\begin{equation}
ds^2=2{\frac{r^2}{P^2}}d\zeta d\bar\zeta-2dudr-\Big[\Delta\ln P-2r(\ln P)_{,u}-
       {\frac{2m}{r}}-{\frac\Lambda3}r^2\Big]du^2\ , \label{E4.25}
\end{equation}
where $\zeta$ is a complex spatial coordinate, $r$ is an affine parameter
along the rays generated by the null vector field
{\bf k}, $u$ is a retarded time, $m$ is a function of $u$ which in some
cases can be interpreted as mass,
and ${\Delta\equiv2P^2\partial^2/\partial\zeta\partial\bar\zeta}$.
The function ${P\equiv P(\zeta,\bar\zeta,u)}$ satisfies the \mbox{Robinson-Trautman} equation
${\Delta\Delta(\ln P)+12m(\ln P)_{,u}-4m_{,u}=0}$.
In this section we restrict attention to the solutions of type {\it N\ }
and denote these as $RTN(\Lambda)$.
In this case  $m=0$ and $\Delta\ln P= K(u)$.
By a transformation
$u=g(\tilde u)$, $r={\tilde r/\dot g}$,  where $\dot g={dg/ d\tilde u}$,
we can set the Gaussian curvature $K(u)$ of the 2-surfaces
$2P^{-2}{d\zeta d\bar\zeta}$ to be $K=2\epsilon$, where
${\epsilon=+1,0,-1}$.
Thus, the different subclasses  can be denoted as
$RTN(\Lambda,\epsilon)$. The corresponding
metrics can be written as
\begin{equation}
ds^2=2{\frac{r^2}{P^2}}d\zeta d\bar\zeta-2dudr-2\Big[\epsilon-r(\ln P)_{,u}
       -{\frac\Lambda6}r^2\Big]du^2\ .\label{E4.29}
\end{equation}
Since ${\epsilon=+1,0,-1}$ and ${\Lambda>0}$, ${\Lambda=0}$,
$\Lambda<0$, 9 invariant subclasses exist.

Another coordinate system for the $RTN(\Lambda,  \epsilon)$
class has been given.
The metric is expressed in terms of a function $f(\xi,u)$ which is an
arbitrary function of $u$, analytic in spatial coordinate $\xi$:
\begin{equation}
ds^2=2v^2d\xi d\bar\xi+2v\bar {\Aa} d\xi du+2v\Aa d\bar\xi du
 + 2 \psi dudv+2(\Aa\bar {\Aa}+\psi \BB)du^2\ , \label{E4.30}
\end{equation}
where
$$\Aa=\epsilon\xi-v f\ ,\qquad
\BB=-\epsilon+{\frac{v}2}(f_{,\xi}+\bar f_{,\bar\xi})+\la
     v^2\psi\ ,\qquad
\psi=1+\epsilon\xi\bar\xi\ .$$
The non-vanishing Weyl tensor components are proportional to $f_{,\xi\xi\xi}$
so that the solutions are conformally flat if $f$ is quadratic in $\xi$.
Thus, the $RTN(\Lambda, \epsilon)$ solutions (\ref{E4.30}) with
$f=f_c= c_0(u)+c_1(u)\xi+c_2(u)\xi^2$, where $c_i(u)$ are arbitrary
complex functions of $u$, are isometric to
Minkowski (if $\Lambda=0$), de~Sitter ($\Lambda>0$)  and
anti-de~Sitter ($\Lambda<0$) spacetime.

It is natural to base the local characterization of these spacetimes on the
equation of geodesic deviation
\begin{equation}
\frac{D^2Z^\mu}{d\tau^2}= -R^\mu_{\ \alpha\beta\gamma}u^\alpha Z^\beta
u^\gamma\ , \label{E2.3}
\end{equation}
where $\uu={{d\bx}/{d \tau }}$, $\uu\cdot\uu=-1$, is the four-velocity of a
free test particle (observer), $\tau$ is the proper time, and $\Z (\tau)$ is
the displacement vector. In order to obtain  invariant results, one sets up a
frame $\{ \e_{(a)} \}$ along the geodesic. Choosing
$\e_{(0)}=\uu$ and perpendicular space-like unit vectors
$\{\e_{(1)},\e_{(2)},\e_{(3)}\}$ in the local hypersurface orthogonal to $\uu$
and by projecting
(\ref{E2.3}) onto the frame, we get
\begin{equation}
\ddot Z^{(i)}=-R^{(i)}_{\ (0)(j)(0)}Z^{(j)}\ , \label{E2.5a}
\end{equation}
where $Z^{(j)}=\e^{(j)}\cdot\Z=e^{(j)}_\mu Z^\mu$ determine directly the
distance between close test particles,
\begin{equation}
\ddot Z^{(i)}\equiv\e^{(i)}\cdot{\frac{D^2\Z}{d\tau^2}}
  =e^{(i)}_\mu{\frac{D^2Z^\mu}{d\tau^2}} \label{E2.5b}
\end{equation}
are physical relative accelerations, and
${R_{(i)(0)(j)(0)}=e^\alpha_{(i)}u^\beta e^\gamma_{(j)}
 u^\delta R_{\alpha\beta\gamma\delta}}$.
Setting $Z^{(0)}=0$,
all test particles are ``synchronized'' by $\tau$.
In the $KN(\Lambda)$ and $RTN(\Lambda,\epsilon)$
spacetimes the equations of geodesic deviation take the form
\begin{eqnarray}
\ddot Z^{(1)}&=&{\frac\Lambda3}Z^{(1)}-\A_+ Z^{(1)}+\A_\times Z^{(2)}
   \ ,\nonumber\\
\ddot Z^{(2)}&=&{\frac\Lambda3}Z^{(2)}+\A_+Z^{(2)}+\A_\times Z^{(1)}
   \ ,\label{E4.12}\\
\ddot Z^{(3)}&=&{\frac\Lambda3}Z^{(3)}\ ,\nonumber
\end{eqnarray}
where the amplitudes of the (exact) gravitational waves are given by
\begin{equation}
\A_+(\tau)={\pul} pq\dot u^2\> \R e\left\{f_{,\xi\xi\xi}\right\}\ ,\qquad
\A_\times (\tau)={\pul} pq\dot u^2\> \I m\left\{f_{,\xi\xi\xi}\right\}
   \ , \label{E4.13}
\end{equation}
for the $KN(\Lambda)$  spacetimes, and by
\begin{equation}
\A_+(\tau)=-{\pul} {\frac{\psi}v}\dot u^2\> \R e\left\{f_{,\xi\xi\xi}\right\}
  \ ,\qquad
\A_\times(\tau)=-{\pul} {\frac{\psi}v}\dot u^2\> \I
m\left\{f_{,\xi\xi\xi}\right\}
   \ , \label{E6.12}
\end{equation}
in the $RTN(\Lambda,\epsilon)$ spacetimes.
Equations (\ref{E4.12})-(\ref{E6.12}) give relative accelerations
of the free test particles in terms of their actual positions. They enable us
to draw a number of simple conclusions:
\begin{enumerate}
\item  All particles move isotropically one with respect to the other
if no gravitational wave is present, i.e., if
$f_{,\xi\xi\xi}=0$. In this case both the $KN(\Lambda)$ and
$RTN(\Lambda,\epsilon)$ spacetimes are vacuum conformally flat,
and therefore Minkowski ($\Lambda=0$), de Sitter ($\Lambda>0$)
and anti-de Sitter ($\Lambda<0$). Such spaces
are maximally symmetric, homogeneous, isotropic, and they represent a natural
background for other ``non-trivial'' $KN(\Lambda)$ and $RTN(\Lambda,\epsilon)$
type {\it N} solutions.
\item If amplitudes $\A_+$ and $\A_\times$ do not vanish
($f_{,\xi\xi\xi}\not=0$), the particles are influenced by the wave
in a similar way as they are affected
by a standard gravitational wave on Minkowski background.
However, if $\Lambda\not=0$, the influence of the  wave adds with the (anti-)
de Sitter isotropic expansion (contraction). This makes plausible our
interpretation of the $KN(\Lambda)$ and $RTN(\Lambda,\epsilon)$ metrics as {\it
exact gravitational waves propagating  on the constant curvature backgrounds}.
\item The wave propagates in the space-like direction of $\e_{(3)}$ and has a
{\it transverse character} since only motions in the perpendicular directions
of $\e_{(1)}$ and $\e_{(2)}$ are affected. The propagation direction given by
$\e_{(3)}$ coincides with the projection of the Debever-Penrose vector $\bk$ on
the hypersurface orthogonal to the observer's velocity $\uu$.
\item There are {\it two polarization modes} of the wave:
``+''  and ``$\times$'', $\A_+$ and $\A_\times$ being the
amplitudes. Under rotation in the transverse plane they transform
in such a way so that the helicity of the wave is 2, as with
linearized waves on Minkowski background.
\item The waves have amplitude $\A=\pul pq\dot u^2\>
\left|f_{,\xi\xi\xi}\right|$ for the $KN(\Lambda)$ class and $\A=\pul (\psi/
v)\dot u^2\> \left|f_{,\xi\xi\xi}\right|$ for $RTN(\Lambda,\epsilon)$; this is
invariant under rotations in the transverse plane.
\end{enumerate}

For a \emph{special} class of timelike geodesics characterized by
$\xi=\xi_0=\text{constant}$ one can calculate
the wave amplitudes explicitly  in terms of $\tau$ for both classes.
At large ${\tau>0}$ the amplitudes decay as ${\A\sim
\exp(-n{\sqrt{\Lambda/3}}\,\tau)}$, ${n>0}$, i.e., {\it waves are damped exponentially}.
The spacetimes approach asymptotically the de~Sitter universe. This is an explicit
demonstration of the {\it cosmic no-hair  conjecture} (see, e.g., Ref.~\refcite{Maeda:1989})
under the presence of waves within exact model spacetimes. In the following section
we turn to the cosmic no-hair
conjecture in the Robinson-Trautman spacetimes of Petrov type {\it II}.

Here, let us remark yet that a suitable physical interpretation of
exact radiative spacetimes which are not asymptotically flat has been achieved
by using the equation of geodesic deviation also
in other cases. For example, Siklos solutions have been shown to represent exact
gravitational waves in the anti-de~Sitter universe.\cite{Podolsky:1998}

\section{Robinson--Trautman radiative spacetimes with ${\Lambda\neq0}$}
\label{sec:rtwl}

Robinson-Trautman metrics are the general radiative vacuum solutions which
admit a geodesic, shearfree and twistfree null congruence of diverging rays. In
the standard coordinates the metric has the form given by Eq.~\eqref{E4.25},
where  ${P\equiv P (u, \zeta, \bar {\zeta})}$ satisfies
the Robinson-Trautman fourth-order parabolic equation
${\Delta \Delta (ln P) + 12 m (ln P), _u - 4 m, _u = 0.}$
In the previous section we considered the Robinson-Trautman metrics of Petrov type {\it N}
when the mass function $m=0$. However, the best candidates
for describing radiation from isolated sources are the
Robinson-Trautman metrics of type {\it II} with the 2-surfaces $S^2$ (having spherical topology)
given by ${u, r =\text{constant}}$ and ${m\neq0}$. The Gaussian curvature of $S^2$ can be
expressed as $K = \Delta lnP$. If ${K = \text{constant}}$, we obtain the Schwarzschild
solution with mass ${m=K^{-\frac {3}{2}}}$.

In the studies of the Robinson-Trautman spacetimes of Petrov type {\it II} with
${\Lambda=0}$ it was shown that these spacetimes exist globally
for all positive ``times'' and converge asymptotically to a
Schwarzschild metric (see Ref.~\refcite{Chrusciel:1992} and references therein).
Interestingly, the extension of these spacetimes across
the "Schwarz\-schild-like" event horizon can only be made with a finite degree
of smoothness. All these rigorous studies are based on the derivation and
analysis of an asymptotic expansion describing the long-time behavior of the
solutions of the nonlinear parabolic Robinson-Trautman equation given above.

\begin{figure}
\begin{center}
\includegraphics[width=7.5cm]{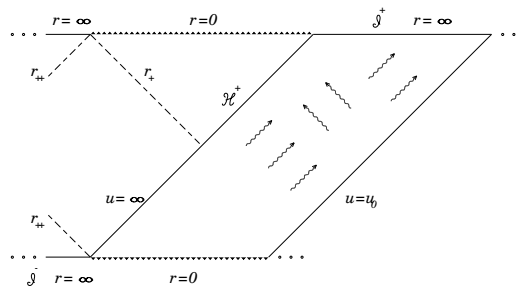}
\caption{\label{fig:rt}
The evolution of the cosmological Robinson-Trautman solutions with a
positive cosmological constant. A black hole with the horizon
$\mathcal H^+$ is formed; at future infinity $\scri^+$ the spacetime approaches
a de Sitter spacetime exponentially fast, in accordance with the cosmic
no-hair conjecture.
}
\end{center}
\end{figure}

We analyzed  Robinson-Trautman radiative spacetimes
with the positive cosmological constant $\Lambda$ in detail.\cite{BicakPodolsky:1995,BicakPodolsky:1997}
The results proving the global
existence and convergence of the solutions of the Robinson-Trautman equation
can be taken over from the previous studies since $\Lambda$ does not
explicitly enter this equation. We have shown that, starting with arbitrary,
smooth initial data at $u=u_0$ (see Fig. \ref{fig:rt}), the cosmological
Robinson-Trautman solutions converge exponentially fast to a Schwarzschild-de
Sitter solution at large retarded times ($u\to \infty$). The interior of a
Schwarzschild-de Sitter black hole can be joined to an "external" cosmological
Robinson-Trautman spacetime across the horizon $\mathcal H^+$.
In some cases this joining can be made with a higher
degree of smoothness than in the corresponding case with $\Lambda = 0$. We have
also demonstrated that the cosmological Robinson-Trautman solutions represent
explicit models exhibiting the cosmic no-hair conjecture: a geodesic observer
outside of the black-hole horizon will see, inside his past light cone, these
spacetimes to approach the de Sitter spacetime exponentially fast as he is
approaching future (spacelike) infinity ${\scri}^+$. For a freely falling
observer the observable universe thus becomes quite bald. This is what the
cosmic no-hair conjecture claims. As far as we are aware, these models
represent the only exact analytic demonstration of the cosmic no-hair
conjecture under the presence of gravitational waves. They also appear to be
the only exact examples of a black-hole formation in nonspherical spacetimes
which are not asymptotically flat. Hopefully, these models may serve as tests
of various approximation methods, and as test beds in numerical studies of more
realistic situations in cosmology.

\section{Fields of uniformly accelerated sources in de~Sitter spacetime}
\label{sec:fieldsinds}

The question of the electromagnetic field and associated radiation from
uniformly accelerated charges has been one of the best known \vague{perpetual
problems} in classical physics from the beginning of the last century. In the
pioneering work in 1909, Born gave the time-symmetric solution for the field of
two point particles with opposite charges, uniformly accelerated in opposite
directions in Minkowski space (cf.\ Fig.~\ref{fig:brst}).
Even the December 2000 issue of
Annals of Physics contains three papers\cite{EriksenGron:2000} with numerous
references on \vague{electrodynamics of hyperbolically accelerated charges}.

In general relativity, solutions of Einstein's equations, representing
\vague{uniformly accelerated particles or black holes}, are the \emph{only}
explicitly known exact \emph{radiative} spacetimes describing \emph{finite}
sources. They were briefly described in Sec.~\ref{sec:boostrot}.

Here, we present the summary of our recent work\cite{BicakKrtous:FUACS}
in which we generalized the Born solutions for scalar
and electromagnetic fields to the case of two charges uniformly accelerated in
de~Sitter universe and explicitly have shown how in the limit ${\Lambda\rightarrow0}$
the Born solutions are retrieved. We also studied the asymptotic expansions of
the fields in the neighborhood of future infinity $\scri^+$. Since in de~Sitter
spacetime conformal infinities $\scri^\pm$ are spacelike, there exist
particle and event horizons. It is known\cite{PenroseRindler:book}
that the radiation field is \vague{less invariantly}
defined when $\scri^+$ is spacelike (it depends on the direction in which
$\scri^+$ is approached), but no explicit model appears to be available so far.
Our solutions can serve as prototypes for studying these issues.

\begin{figure}
\begin{center}
\includegraphics{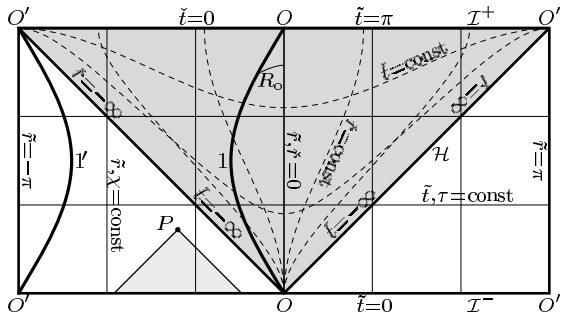}
\caption{\label{fig:dS}
The conformal diagram of de~Sitter spacetime. Uniformly accelerated particles
move along worldlines $1$ and $1'$. The shaded region is the domain of
influence of $1$, its boundary $\mathcal{H}$ is the \vague{creation light cone}
of this particle \vague{born} at ${\tlt=0}$ at \vague{point} $O$. Retarded
fields of $1$ and $1'$ cannot affect point $P$, a Coulomb-type field, however,
cannot vanish there.}
\end{center}
\end{figure}

In another work\cite{BicakKrtous:ASDS} we analyzed fields of accelerated
sources to show the \emph{insufficiency of purely retarded fields in de~Sitter
spacetime}. Consider a point $P$ near $\scri^-$ whose past null cone will not
cross the particles' worldlines (Fig.~\ref{fig:dS}). The field at $P$ should
vanish if an incoming field is absent. However, the \vague{Coulomb-type} field
of particles cannot vanish there because of Gauss law\cite{Penrose:1967}. The
requirement that the field be purely retarded leads in general to a bad
behavior of the field along the \vague{creation light cone} of the
\vague{point} at which a source enters the universe (see
Ref.~\refcite{BicakKrtous:ASDS} for a detailed discussion).

The de~Sitter universe has topology ${S^3\times\realn}$. The metric in standard
\vague{spherical} coordinates is
\begin{equation}\label{dSMtrctauchi}
  \mtrc_\dS = -\grad\tau\formsq +
  \alpha^2 \cosh^2(\tau/\alpha)\;
  \bigl(\grad\chi\formsq
  +\sin^2\chi\;\sphmtrc\bigr)\commae
\end{equation}
where
  $\sphmtrc = \grad\tht\formsq + \sin^2\!\tht\,\grad\ph\formsq$,
  ${\tau\in\realn}$, and $\alpha^2={3}/{\Lambda}$.
Putting
  $\chi=\tlr$, $\tau = \alpha\log\tan(\tlt/2)$,
  $\tlt\in\langle0,\pi\rangle$,
in Eq.~\eqref{dSMtrctauchi}, the de~Sitter metric can be written in the form
\begin{equation}\label{dSMtrctl}
  \mtrc_\dS = \alpha^2\sin^{-2}\tlt\;\bigl(
  -\grad\tlt\,\formsq+\grad\tlr\formsq+
  \sin^2\!\tlr\;\sphmtrc\bigr)\period
\end{equation}
The lines ${\tlr=\pi}$ and ${\tlr=-\pi}$ are identified, the spacelike
hypersurfaces ${\tlt=0,\pi}$ represent $\scri^-$ and $\scri^+$
(Fig.~\ref{fig:dS}).

We recently studied\cite{BicakKrtous:FUACS,BicakKrtous:ASDS}
two particles moving with uniform acceleration in
de~Sitter space. Their worldlines are plotted in Fig.~\ref{fig:dS} as $1$, $1'$
(for explicit formulae see Ref.~\refcite{BicakKrtous:ASDS}, Eq.~(4.4)).
Both particles start at
antipodes of the spatial section of de~Sitter space at $\scri^-$ and move one
towards the other until ${\tlt=\pi/2}$, the moment of the maximal contraction
of de~Sitter space. Then they move, in a time-symmetric manner, apart from each
other until they reach future infinity at the antipodes from which they
started. Their physical velocities, as measured in the \vague{comoving}
coordinates
  ${\{\tau,\chi,\tht,\ph\}}$,
have simple form
  ${v_\chi=\sqrt{\mtrc_{\chi\chi}}\,{d\chi}/{d\tau}}=
  {\mp a_\oix\alpha\tanh(\tau/\alpha)\,
  [1+a_\oix^{2}\alpha^2\tanh^2(\tau/\alpha)]^{-1/2}}$,
where $\abs{a_\oix}$ is the magnitude of their acceleration. In contrast to the
flat space case, the particles do \emph{not} approach the velocity of light in the
\vague{natural} global coordinate system. Nevertheless, they are causally disconnected
(Fig.~\ref{fig:dS}) as in the flat space case: no signal from one particle can
reach the other particle.

Two charges moving along the orbits of the boost Killing vector in flat space
are \emph{at rest} in the Rindler coordinate system and have a constant
distance from the spacetime origin, as measured along the slices orthogonal to
the Killing vector. Similarly, the worldlines $1$ and $1'$ are the orbits of
the \vague{static} Killing vector ${\cvectil{T}}$ of de~Sitter space. In static
coordinates
  ${\{T,R,\tht,\ph\}}$,
  $T = {\frac\alpha2 \log[({\cos\tlr\mspace{-0.5mu}-\mspace{-0.5mu}\cos\tlt\mspace{1.5mu}})/({\cos\tlr\mspace{-0.5mu}+\mspace{-0.5mu}\cos\tlt\mspace{1.5mu}}})]$,
  $R = {\alpha\, {\sin\tlr}/{\sin\tlt}}$,
the particles $1$, $1'$ are at rest at
  ${R \mspace{-2mu}=\mspace{-2.5mu} \pm R_\oix \mspace{-2.5mu}=\mspace{-2.5mu} \mp{a_\oix\alpha^2}/{\sqrt{1\!+\!a_\oix^2\alpha^2}}}$,
with four-accelerations $-(R_\oix/\alpha^2)\,\cvectil{R}$. The particle $1$
($1'$) has, as measured at fixed $T$, a constant proper distance from the
origin ${\tlt=\pi/2}$, ${\tlr=0}$ (${\tlr=\pi}$). As with Rindler coordinates
in Minkowski space, the static coordinates cover only a \vague{half} of
de~Sitter space; in the other half the Killing vector $\cvectil{T}$ becomes
spacelike.

By the conformal transformation of the boosted Cou\-lomb fields in Minkowski
space, we constructed~\cite{BicakKrtous:ASDS} test scalar and electromagnetic
fields produced by charges moving along the worldlines $1$, $1'$ in de~Sitter
space. The scalar field from two \emph{identical} scalar charges $\SFq$ is
given~by
\begin{gather}
  \SF_\sym = (\SFq/{4\pi})\;\Kfact^{-1}
  \commae\label{SFsym}\displaybreak[0]\\
  \Kfact =
  \bigl[\alpha^2\bigl(\sqrt{1+a_\oix^{2}\alpha^2}
  + a_\oix R \cos\tht\bigr)^2
  - \alpha^2+R^2\bigr]^{\frac12}
  \label{Kfact}
\end{gather}
(Ref.~\refcite{BicakKrtous:ASDS}, Eq.~(5.4)), whereas the electromagnetic field
due to \emph{opposite} charges $+\EMq$ and $-\EMq$ is
(Ref.~\refcite{BicakKrtous:ASDS}, Eq.~(5.7))
\begin{equation}\label{EMsymtl}
\begin{split}
  \EMF_\sym &= -\frac{\EMq}{4\pi}\,
  \frac1{\Kfact^3}
  \frac{a_\oix\,\alpha^4}{\sin^3\tlt}
  \;\Bigl[
  \cos\tlt\sin^2\tlr\sin\tht\,
  \grad\tlr\wedge\grad\tht
  \\&\mspace{12mu}
  +(a_\oix^{-1}\sqrt{a_\oix^{2}+\alpha^{-2}}\sin\tlr +
  \sin\tlt\cos\tht)\,
  \grad\tlt\wedge\grad\tlr
  \\&\mspace{12mu}
  -\sin\tlt\cos\tlr\sin\tlr\sin\tht\,
  \grad\tlt\wedge\grad\tht
  \Bigr]\period
\end{split}
\end{equation}
\begin{figure}
\begin{center}
\includegraphics{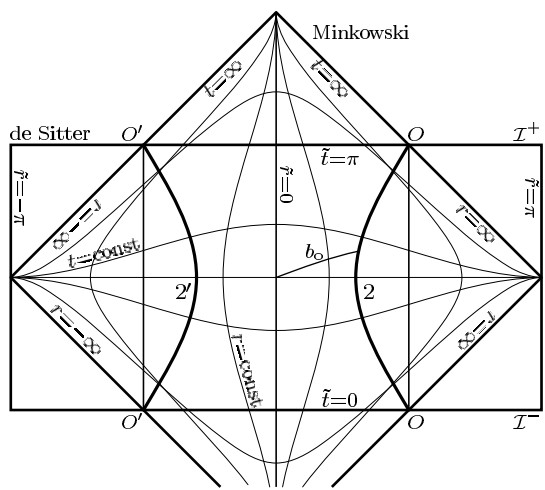}
\caption{\label{fig:dSMink}
The worldlines $2$, $2'$ of uniformly accelerated charges symmetrically located
with respect to the origins of both de~Sitter and conformally related Minkowski
spacetimes.}
\end{center}
\end{figure}
We call these smooth (outside the sources) fields symmetric because they can be
written as a symmetric combination of retarded and advanced effects from both
charges.

Although Eqs. \eqref{SFsym} and \eqref{EMsymtl} represent fields due to
uniformly accelerated charges in de~Sitter space, their relation to the Born
solutions is not transparent because the sources are not located symmetrically
with respect to ${\tlr=0}$. Hence, we considered the worldlines $2$ and $2'$
(Fig.~\ref{fig:dSMink}) which, due to homogeneity and isotropy of de~Sitter
space, also represent uniformly accelerated particles. These worldlines and the
resulting fields can be obtained from Eqs.~\eqref{SFsym}--\eqref{EMsymtl} by a
spatial rotation by $\pi/2$. We find the worldlines $2$, $2'$ to be given by
${\cot\tlt = -{\sinh\bigl(\lambda_\dS \alpha^{-1}
  \sqrt{1+a_\oix^2\alpha^2}\bigr)}/\sqrt{1+a_\oix^2\alpha^2}}$,
${\tan\tlr = \pm{\cosh\bigl(\lambda_\dS \alpha^{-1}
  \sqrt{1+a_\oix^2\alpha^2}\bigr)}/({a_\oix\alpha})}$,
${\tht=0}$, ${\ph=0}$.
The scalar and electromagnetic fields are
\begin{gather}
  \SF_\dSBorn = ({\SFq}/{4\pi})\;
  {\sin\tlt}\; ({\sin\tlt+\cos\tlr})^{-1}\;
  {\retR}^{-1}\commae\label{SFdSBorn}\displaybreak[0]\\
\begin{split}
  \EMF_{\dSBorn} &= -\frac{\EMq}{4\pi}
  \frac{\alpha^3}{\retR^3}
  \frac{a_\oix\alpha\sin\tht}{(\sin\tlt+\cos\tlr)^3}\bigl[
  \sin^2\tlr\cos\tlt\,
  \grad\tlr\wedge\grad\tht
  \\&\mspace{2mu}
  - (a_\oix^{-1}\sqrt{a_\oix^{2}+\alpha^{-2}} \cos\tlr - \sin\tlt\mspace{2mu})\,
  \cot\tht\,\grad\tlt\wedge\grad\tlr
  \\&\mspace{2mu}
  + (a_\oix^{-1}\sqrt{a_\oix^{2}+ \alpha^{-2}} - \cos\tlr\sin\tlt\mspace{2mu})\,
  \sin\tlr\,\grad\tlt\wedge\grad\tht
  \bigr]\commae\label{EMFdSBorn}
\end{split}\displaybreak[0]\\
  \frac{\retR}{\alpha} = \frac{
  \bigl[(a_\oix\alpha\sin\tlt -
  \sqrt{1+a_\oix^2 \alpha^2}\cos\tlr)^2 +
  \sin^2\tlr\sin^2\tht\bigr]^{\frac12}}
  {\sin\tlt+\cos\tlr}\period\label{RetR}\nonumber
\end{gather}

In order to understand explicitly the relation of these fields to the classical
Born solutions, we considered Minkowski spacetime with spherical coordinates
${\{\nlt, \nlr, \tht, \ph\}}$ and set
${\nlt = {-\alpha \cos\tlt}/(\cos\tlr+\sin\tlt\,)}$,
${\nlr = {\alpha \sin\tlr}/(\cos\tlr+\sin\tlt\,)}$.
In coordinates ${\{\nlt,\nlr,\tht,\ph\}}$, which can also be used in de~Sitter
space, the worldlines $2$, $2'$
acquire the simple form: ${\tht=0}$, ${\ph=0}$, and
${\nlt = b_\oix \sinh({\lambda_\nlMink}/{b_\oix})}$,
${\nlr = \pm b_\oix \cosh({\lambda_\nlMink}/{b_\oix})}$,
where $\lambda_\nlMink$ is the proper time as measured by Minkowski metric, and
${b_\oix}/{\alpha} \!=\! \sqrt{1\!+\!a_\oix^2\alpha^2}-a_\oix\alpha$. These
worldlines are just two hyperbolae (Fig.~\ref{fig:dSMink}),
representing particles with uniform acceleration $1/b_\oix$ as measured in
Minkowski space.
Transforming the fields \eqref{SFdSBorn} and \eqref{EMFdSBorn} into conformally
flat coordinates ${\{\nlt, \nlr, \tht, \ph\}}$, we obtain
the fields in the form which in the limit ${\Lambda\to0}$ yields easily the
classical Born fields of uniformly accelerated charges in Minkowski space
(see Ref.~\refcite{BicakKrtous:FUACS} for details).

What is the character of the generalized Born fields? Focusing on the
electromagnetic case, we first decompose the field \eqref{EMFdSBorn} into the
orthonormal tetrad  tied to coordinates
${\{\tlt,\tlr,\tht,\ph\}}$. Splitting the field into the
electric and magnetic parts we get
\begin{equation}\label{EMEBcb}
\begin{aligned}
  \EME&=\frac{\EMq}{4\pi}
  \frac{\alpha\sin^2\tlt}{\retR^3(\sin\tlt+\cos\tlr)^3}
  \times\\&\qquad
  \bigl[
  -\bigl(\sqrt{1+a_\oix^2\alpha^2}\cos\tlr
  -a_\oix\alpha\sin\tlt\bigr)\cos\tht\,\cbv_\tlr
  \\&\qquad
  +\bigl(\sqrt{1+a_\oix^2\alpha^2}
  -a_\oix\alpha\sin\tlt\cos\tlr\bigr)\sin\tht\,\cbv_\tht
  \bigr]\commae\\
  \EMB&=-\frac{\EMq}{4\pi}
  \frac{a_\oix\alpha^2\sin^2\tlt}
  {\retR^3(\sin\tlt+\cos\tlr)^3}
  \cos\tlt\sin\tlr\sin\tht\,\cbv_\ph\period
\end{aligned}
\end{equation}
The fields exhibit some features typical for the classical Born solution. The
toroidal electric field, $\EME_\ph$, vanishes, only $\EMB_\ph$ is
non-vanishing. At $\tlt=\pi/2$, the moment of time symmetry, $\EMB_\ph=0$. It
vanishes also for $\tht=0$ --- there is no Poynting flux along the axis of
symmetry.

The classical Born field decays rapidly (${\EME\sim r^{-4}}$, ${\EMB\sim
r^{-5}}$) at spatial infinity, but it is \vague{radiative} (${\EME,\EMB}\sim
r^{-1}$) if we expand it along null geodesics ${t-r=\text{constant}}$,
approaching thus null infinity. (This behavior is typical also for
boost-rotation symmetric spacetimes with ${\Lambda=0}$ mentioned in Sec.~\ref{sec:boostrot}.)
In de~Sitter spacetime with standard slicing,
the space is finite ($S^3$). However, we can approach infinity along spacelike
hypersurfaces if, for example, we consider the \vague{steady-state} half of
de~Sitter universe (cf. Fig.~\ref{fig:dS}) with flat-space slices, i.e., if we
take the \vague{conformally flat} time ${\hct=\text{constant}}$;
the tetrad components of the fields then
decay as $\hcr^{-2}$.

The fields decay very rapidly along \emph{timelike} worldlines as $\scri^+$ is
approached. This is caused by the exponential expansion of slices
${\tau=\text{constant}}$ (cf. Eq.~\eqref{dSMtrctauchi}). As
${\tau\rightarrow\infty}$ the electric field \eqref{EMEBcb} becomes radial,
${\EME_\tlr\sim\exp(-2\tau/\alpha)}$, and ${\EMB_\ph\sim\exp(-2\tau/\alpha)}$.
The energy density, ${\EMU=\frac12(\EME^2+\EMB^2)}$, decays as
${(\text{expansion factor})^{-4}}$
--- as energy density in the radiation dominated standard cosmologies.
This behavior corresponds to the \vague{cosmic no-hair} properties in
spacetimes with $\Lambda>0$ discussed in Secs.~\ref{sec:nwl} and \ref{sec:rtwl}.

\begin{figure}
\begin{center}
\includegraphics{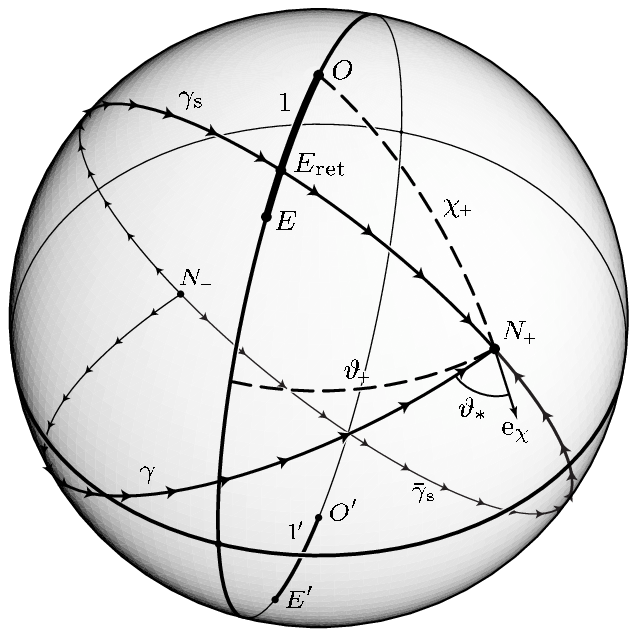}
\caption{\label{fig:spacesec}
Space trajectories of null geodesics $\gamma$, $\gamma_{\mathrm{s}}$ and
$\bar\gamma_{\mathrm{s}}$ indicated on the slice ${\tlt=\text{constant}}$
(${\ph=0}$). Charges $1$, $1'$ move along $\tht=0$ from poles $O$, $O'$ to
points $E$, $E'$ and back. $\gamma$, $\gamma_{\mathrm{s}}$ and
$\bar\gamma_{\mathrm{s}}$ start at $N_\mix$ at ${\tlt=0}$ and arrive at
$N_\pix$ (with coordinates $\chp$, $\thp$) at $\tlt=\pi$. The direction of
$\gamma$ at $N_\pix$ is specified by angles $\thst$, $\phst$ ($\phst$ describes
rotation around $\cbv_\chi$ in the dimension not seen). $\gamma_{\mathrm{s}}$
crosses the worldline of particle $1$ at $E_{\mathrm{ret}}$,
$\bar\gamma_{\mathrm{s}}$ reaches $N_\pix$ from the opposite direction.}
\end{center}
\end{figure}

To study the asymptotic behavior of a field along a null geodesic,
we have to (i) find a geodesic and
parameterize it by an affine parameter $\afp$, (ii) construct a tetrad
parallelly propagated along the geodesic, and (iii) study the asymptotic
expansion of the tetrad components of the field. We find that along null
geodesics lying in the axis $\tht=0$ (thus crossing the particles' worldlines)
the \vague{radiation field}, i.e. the coefficient of the leading term in
${1/\afp}$, vanishes, as could have been anticipated
--- particles do not radiate in the direction of their acceleration.
The radiation field also vanishes along null geodesics reaching infinity along
directions \emph{opposite} to those of geodesics emanating from the particles
(see Fig.~\ref{fig:spacesec}). Along all other geodesics the field \emph{has}
radiative character. Along a null geodesic coming from a general direction to a
general point on $\scri^+$ we find the electric and magnetic fields (in a
parallelly transported tetrad~${\{\pbv_\mu\}}$) to be perpendicular one to the
other, equal in magnitude, and proportional to $\afp^{-1}$. The magnitude of
Poynting flux,
${\abs{\EMP_{(\pbv)}}}={\abs{\EME_{(\pbv)}}{}^{\!2}}={\abs{\EMB_{(\pbv)}}{}^{\!2}}$,
is
\begin{equation}\label{radpatt}
\begin{split}
  &\abs{\EMP_{(\pbv)}} \!=
  \frac{\EMq^2}{(4\pi)^2}
  \frac{a_\oix^2\sin^2\thp\csc^4\chp}{4(1+a_\oix^{2}\alpha^2\cos^2\thp)^{3}}
  \bigl[
  \cos^2\thst\sin^2\phst
  \\&\mspace{64mu}
  +\bigl(\cos\phst +
  a_\oix^{-1}\sqrt{a_\oix^{2}+\alpha^{-2}}
  \sin\thst\csc\thp\bigr)^2
  \bigr]\;\afp^{-2}
\end{split}
\end{equation}
(see Fig.~\ref{fig:spacesec} for the definition of angles $\chp$, $\thp$,
$\thst$, $\phst$). These results are typical for a \emph{radiative} field. Most
interestingly, this radiative aspect depends on the specific geodesic along
which a given point on \emph{spacelike} $\scri^+$ is approached
(cf.~\cite{PenroseRindler:book}). Moreover, the radiative character does not
disappear even for static sources but it does along null geodesics emanating
from such sources.

Summarizing, we have analyzed the fields of uniformly accelerated charges in
a de Sitter universe which go over to classical Born's fields in the limit
${\Lambda\rightarrow 0}$. Aside from some similarities found, the generalized
fields provide the models showing how a positive cosmological constant implies
essential differences from physics in flat spacetime: advanced effects occur
inevitably, and the character of the far fields depends substantially on the
way in which future (spacelike) infinity is approached.

\section{Outlook}
\label{sec:outlook}

One of the best known boost-rotation symmetric spacetimes is the vacuum
C-metric, representing two black holes uniformly accelerated in opposite
directions due to ``strings'' located either between them, or extending
from each of them to infinity. This solution can be generalized
to include charged and rotating black holes, and also a non-zero
cosmological constant. The ``cosmological'' C-metric has attracted some
attention not long ago (see, e.g., Refs.~\cite{MannRoss:1995,DiasLemos:2003b})
but its radiative properties have not been investigated until very
recently\cite{KrtousPodolsky:RABHDS}. Notice that two objects uniformly
accelerated in opposite directions in de~Sitter-like universe do not
approach velocity of light asymptotically and form a permanently
bounded system --- in contrast to analogous objects in spacetimes with
${\Lambda=0}$ (compare Figs.~\ref{fig:brst} and \ref{fig:dS}, \ref{fig:dSMink}).
By a detailed analysis of the asymptotic behavior of the ``charged, cosmological''
C-metric near future spacelike infinity, the peeling properties
have been demonstrated and the dependence of both gravitational and
electromagnetic fields on spatial directions from which
a point at null infinity is approached exhibited.\cite{KrtousPodolsky:RABHDS} This interesting
effect, typical for spacetimes with a spacelike future infinity,
can thus be explicitly seen here for the first time.
The radiation pattern at a point of $\scri^+$ in the case of
electromagnetic fields due to the exact charged C-metric
is the same as that from test electromagnetic fields from
uniformly accelerated charges in de~Sitter space discussed in
Sec.~\ref{sec:fieldsinds} (see, e.g., Eq.~\eqref{radpatt}).

More generally, it can be demonstrated that the dependence of radiative
parts of the fields on a direction along which a spacelike $\scri^+$
is approached is completely determined by the algebraic (Petrov) type
of the fields.\cite{KrtousPodolskyBicak:GEFDSI}

\section*{Acknowledgments}

We are pleased to thank Ji\v{r}\'{\i} Podolsk\'{y} for
the stimulating collaboration on several topics included in this review.
We have been supported by the grant GA\v{C}R 202/99/026 of the Czech Republic.
J.~B. thanks the organizers of the 10th Greek Relativity Meeting,
in particular Kostas Kokkotas,
for the invitation to Chalkidiki and kind hospitality.

\end{document}